\documentclass[11pt,superscriptaddress,aps,prd]{revtex4}

\everymath{\displaystyle}
\usepackage{graphicx,amsmath,amssymb,amsfonts,amstext,latexsym,stmaryrd,amscd,txfonts,pxfonts,hyperref}
\hypersetup{
    colorlinks=true,
    linkcolor=blue,
    filecolor=magenta,      
    urlcolor=black,
		citecolor=red,
		}

\begin{document}

\title{On the Quantization of Burgers Vector and Gravitational Energy in the Space-Time of a Conical Defect}

\author{\href{https://scholar.google.com/citations?hl=en&user=I_4vtOEAAAAJ}{F. L. Carneiro}}
\email{fernandolessa45@gmail.com}
\affiliation{Instituto de F\'isica, Universidade de Bras\'ilia, 70910-900, Bras\'ilia, DF, Brazil}

\author{\href{https://scholar.google.com/citations?user=DBegoecAAAAJ&hl=en}{S. C. Ulhoa}}
\email{sc.ulhoa@gmail.com}
\affiliation{Instituto de F\'isica, Universidade de Bras\'ilia, 70910-900, Bras\'ilia, DF, Brazil}
\affiliation{International Center of Physics, Instituto de F\'{\i}sica, Universidade de
Bras\'{\i}lia, 70.910-900, Brasilia, DF, Brazil}
\affiliation{Canadian Quantum Research Center, 204-3002 32 Ave Vernon, BC V1T 2L7  Canada}

\author{J. F. da Rocha-Neto}
\email{rocha@fis.unb.br}
\affiliation{Instituto de F\'isica, Universidade de Bras\'ilia, 70910-900, Bras\'ilia, DF, Brazil}

\author{\href{https://scholar.google.com/citations?hl=en&user=F0UBVgUAAAAJ}{J. W. Maluf}}
\email{jwmaluf@gmail.com}
\affiliation{Instituto de F\'isica, Universidade de Bras\'ilia, 70910-900, Bras\'ilia, DF, Brazil}

\begin{abstract}

A conical topological defect is the result of translational and/or rotational deformations of spacetime, in particular the Burgers vector describes the translational deformation. Such a configuration represents a discontinuity, that cannot be removed by coordinate transformations, and is related to the spacetime torsion. Using the Teleparallel Equivalent of General Relativity (TERG), a gravitational theory that is dynamically equivalent to General Relativity (GR), we investigate the consequences of assuming a discrete Burgers vector on the geodesic motion of particles around a static conical defect. The result is a  helical geodesic motion of a test particle around the defect, with a discrete step that depends on the magnitude of the dislocation.
\end{abstract}
\keywords{topological defects, Teleparallel Gravity, discrete Burgers vector.}

\maketitle

\date{\today}

\section{Introduction}

Defects in crystals and metals may occur in three forms: point, linear or planar defects. The linear defects are caused by translational (dislocation) or rotational (disclination) distortions. It was soon realized that such defects in a lattice could be  applied to gravitation. For instance, cosmic strings can be interpreted as topological defects in spacetime \cite{holz1992topological,gal1993spinning,edelen1994space}. The more general spacetime defects solutions were constructed using the Volterra process \cite{volterra1907equilibre} by Puntingam and Soleng \cite{puntigam1997volterra}.

Linear defects in crystals occur when an irregularity in the arrangement of their atoms is present. This feature takes place in a discrete lattice. However, in a continuous limit the torsion is responsible for such linear defects \cite{bilby1955continuous,kondo1952proceedings}. In the context of the General Relativity (GR), a spacetime containing a defect is locally flat, but not topologically flat, i.e., the curvature tensor is null everywhere except on the defect itself. Therefore alternative theories of gravitation that take into account the torsion tensor become relevant to the comprehension of topological defects in spacetime \cite{letelier1995spacetime}. Two of such theories are Einstein-Cartan (EC) theory \cite{soleng1990spin,soleng1992spinning} and the Teleparallel Equivalent of General Relativity (TEGR) \cite{maluf2001space}. In the first one both curvature and torsion are non-vanishing quantities that plays an important role in the dynamics of spacetime. Actually in such an approach the torsion is a consequence of the spin of the matter field. On the other hand, in the framework of TEGR the torsion is entirely responsible by the dynamics of the spacetime, thus the vanishing of torsion implies Minkowski's spacetime. The main advantage of TEGR is the existence of a well defined expression for the gravitational energy-momentum tensor \cite{maluf2013teleparallel}. This former alternative gravitational theory seems to be more suitable to study a spacetime dislocation  which is directly associated to the torsion tensor. It should be noted that the torsion is non-null even outside the defect, thus an experimental approach could reveal the true origin of the gravitational field. In this sense the quantization of Burgers vector has been used as an attempt to introduce the Planck's constant into the structure of spacetime \cite{magnon1991spin,sabbata1991quantum,ross1989planck}. In fact the Plank length $\left(\frac{\hbar G}{c^{3}}\right)^{1/2}$ was understood as a fundamental scale of the defect. This geometric assumption led to a quantization of the intrinsic angular momentum in the EC theory \cite{ross1989planck}. In this paper, this same assumption is made which leads to the possibility of a discrete gravitational energy.

This paper is divided as follows.
In section \ref{terg} we present the TERG and briefly discuss the tetrad fields, explicitly writing the gravitational energy-momentum vector from the Lagrangian formulation of the theory. 
In section \ref{conical} we present the conical defect that will be considered in this paper and the vector that describes the defect intensity and distortion direction, called Burgers vector in solid state. Then we calculate the total energy of the conical defect considered (contains dislocation and disclination) and compare it with the energy of pure dislocations and pure disclinations. The geodesic equations for a massive particle in the presence of the conical defect are then obtained.
In section \ref{results} we pursue the main goal of this article, i.e., to obtain a discrete observational parameter for the particles in the presence of a discrete defect. Also in this section, we obtain a discrete relation for the variation of the classical kinetic energy of the particle.
Finally, in section \ref{conclusions} we present our final comments regarding the results obtained and the possibility of the existence of a discrete gravitational energy for the defect.

Notation: Greek letters $\mu, \nu, \ldots$ denote spacetime indices and Latin letters $a,b,\ldots$ denote SO(3,1) indices, both running from $0$ to $3$. Time is indicated by the spacetime indice $0$ and local indice $(0)$ and space by the spacetime indices $i=1,2,3$ and local indices $(i)=(1),(2),(3)$. The tetrad field is indicated by $e^{a}\,_{\mu}$. Minkowski's flat metric tensor $\eta_{ab}$ raises and lowers the local indices and the metric tensor $g_{\mu\nu}$ raises and lowers the spacetime indices. The geometrized unit system $G=c=1$ is utilized unless otherwise stated.

\section{Teleparallelism Equivalent to General Relativity (TEGR)}\label{terg}

Teleparallelism  Equivalent to General Relativity is a gravitational theory in which the fundamental variables are the tetrad fields $e^{a}\,_{\mu}$. This description is dynamically equivalent to GR but allows the definition of both a gravitational energy-momentum vector and a gravitational energy-momentum tensor.

Tetrads are reference frames adapted to preferred observers in spacetime, with the components $e_{(0)}\,^{\mu}$ always tangent to the observer's worldline. Therefore, in the observer's rest frame, the components $e_{(0)}\,^{\mu}$ may be identified with his four-velocity $u^{\mu}$.
In the TEGR, tetrad fields are auto-parallel, i.e.,
\begin{equation}\label{eq1}
\nabla_{\nu}e^{a}\,_{\mu}=\partial_{\nu}e^{a}\,_{\mu}+\Gamma^{\lambda}\,_{\mu\nu}e^{a}\,_{\lambda}=0\,.
\end{equation}
Isolating the connection in the above equation, we arrive at
\begin{equation}\label{eq2}
\Gamma^{\lambda}\,_{\mu\nu}=e_{a}\,^{\lambda}\partial_{\nu}e^{a}\,_{\mu}\,.
\end{equation}
The above connection is the Weitenzenb\"ock connection and, unlike the Christoffel symbols, it is antisymmetric in the permutation of the last two indices. Thus yield a non-null torsion tensor
\begin{equation}\label{eq3}
T^{\lambda}\,_{\mu\nu}=\Gamma^{\lambda}\,_{\mu\nu}-\Gamma^{\lambda}\,_{\nu\mu}\,,
\end{equation}
that can be written in terms of tetrads as
\begin{equation}\label{eq4}
T^{a}\,_{\mu\nu}=e^{a}\,_{\lambda}T^{\lambda}\,_{\mu\nu}=\partial_{\mu}e^{a}\,_{\nu}-\partial_{\nu}e^{a}\,_{\mu}\,.
\end{equation}
The Weitenzenb\"ock connection is related identically to Christoffel symbols
${}^0{\Gamma}_{\mu \lambda\nu}$ by
\begin{equation}
\Gamma_{\mu \lambda\nu}= {}^0{\Gamma}_{\mu \lambda\nu}+ K_{\mu
\lambda\nu}\,, \label{2}
\end{equation}
where $K_{\mu \lambda\nu}$ is the contortion tensor which is defined by
\begin{eqnarray}
K_{\mu\lambda\nu}&=&\frac{1}{2}(T_{\lambda\mu\nu}+T_{\nu\lambda\mu}+T_{\mu\lambda\nu})\,.\label{3}
\end{eqnarray}
Then using this identity, the scalar of curvature of the Riemann space is written in terms of torsion in Weitzenb\"ock space as
\begin{equation}
eR(e)\equiv -e\left({1\over 4}T^{abc}T_{abc}+{1\over2}T^{abc}T_{bac}-T^aT_a\right)+2\partial_\mu(eT^\mu)\,.\label{5}
\end{equation}
Therefore the Lagrangian density in the TEGR is constructed from a quadratic combination of the torsion tensor
\begin{equation}\label{eq5}
\mathcal{L}=-ke\Sigma^{abc}T_{abc}-\mathcal{L}_{M}\,,
\end{equation}
where $e\equiv det(e^{a}\,_{\mu})$, $k=\frac{1}{16\pi}$, $\mathcal{L}_{M}$ is the Lagrangian density of the matter fields and
\begin{equation}\label{eq6}
\Sigma^{abc}\equiv\frac{1}{4}\left(T^{abc}+T^{bac}-T^{cab}\right)+\frac{1}{2}\left(\eta^{ac}T^{b}-\eta^{ab}T^{c}\right)\,.
\end{equation}
The torsion trace is defined by $T_{a}\equiv T^{b}\,_{ba}$. By varying the Lagrangian density (\ref{eq5}) relative to the tetrad field $e^{a}\,_{\mu}$, we arrive at the field equations \cite{maluf2013teleparallel}
\begin{equation}\label{eq7}
e_{a\lambda}e_{b\mu}\partial_{\nu}\left(e\Sigma^{b\lambda\nu}\right)-e\left(\Sigma^{b\nu}\,_{a}T_{b\nu\mu}-\frac{1}{4}e_{a\mu}T_{bcd}\Sigma^{bcd}\right)=\frac{1}{4k}eT_{a\mu}\,,
\end{equation}
where $eT_{a\mu}\equiv\frac{\delta\mathcal{L}_{M}}{\delta e^{a\mu}}$. The field equations (\ref{eq7}) are dynamically equivalent to those of the GR, being covariant under arbitrary Lorentz and coordinate transformations. In addition, the Lagrangian density (\ref{eq5}) is not invariant under local Lorentz transformations, only under global Lorentz transformations and coordinate transformations, agreeing with the conditions specified by M$\o$ller \cite{moller1964conservation}. The TEGR has a well established Hamiltonian formulation, with quantities such as the gravitational energy-momentum and angular momentum well defined, and satisfying the Poincar\'e group algebra. For a detailed discussion, see Ref. \cite{maluf2013teleparallel} and references therein.

The expression for the total energy-momentum of the spacetime required in this paper may be obtained directly from the field equations (\ref{eq7}), and this expression is in agreement with those obtained in the Hamiltonian formulation of the theory. In order to obtain the energy-momentum tensor for the gravitational field, we rewrite the field equations (\ref{eq7}) as
\begin{equation}\label{eq8}
\partial_{\nu}\left(e\Sigma^{a\lambda\nu}\right)=\frac{1}{4k}ee^{a}\,_{\mu}\left(t^{\lambda\mu}+T^{\lambda\mu}\right)\,,
\end{equation}
where we define
\begin{equation}\label{eq9}
t^{\lambda\nu}=k\left(4\Sigma^{bc\lambda}T_{bc}\,^{\mu}-g^{\lambda\mu}\Sigma^{bcd}T_{bcd}\right)\,.
\end{equation}
Using the fact that the tensor $\Sigma^{a\mu\nu}$ is antisymmetric in the permutation of the last two indices, differentiating the equation (\ref{eq8}) with respect to $\lambda$ and noticing that the left hand side will be zero, we arrive at
\begin{equation}\label{e110}
\partial_{0}\left[ee^{a}\,_{\mu}\left(t^{0\mu}+T^{0\mu}\right)\right]=-\partial_{i}\left[ee^{a}\,_{\mu}\left(t^{i\mu}+T^{i\mu}\right)\right]\,.
\end{equation}
By integrating both sides of the above equation into the arbitrary three-dimensional volume $V$, we obtain
\begin{equation}\label{eq11}
\frac{d}{dt}\int_{V}{d^{3}x\left[ee^{a}\,_{\mu}\left(t^{0\mu}+T^{0\mu}\right)\right]}=-\oint_{S}{dS_{i}\left[ee^{a}\,_{\mu}\left(t^{i\mu}+T^{i\mu}\right)\right]}\,,
\end{equation}
where the surface $S$ is orthogonal to $\partial_{i}$ and delimits the volume $V$.
Equation (\ref{eq11}) is a continuity equation. Since the tensor $\mathcal{T}^{\mu\nu}$ represents the energy-momentum of the matter-radiation fields, the tensor $t^{\mu\nu}$ represents the energy-momentum of the gravitational field. Thus, the total energy-momentum vector contained in the volume $V$ is given by
\begin{equation}\label{eq12}
P^{a}\equiv\int_{V}{d^{3}xee^{a}\,_{\mu}\left(t^{0\mu}+T^{0\mu}\right)}\,.
\end{equation}
With the aid of the field equations (\ref{eq8}), the above equation may be written as
\begin{equation}\label{eq13}
P^{a}=-4k\int_{V}{d^{3}x\partial_{\mu}\Sigma^{a0\mu}}=-4k\int_{V}{d^{3}x\partial_{i}\Sigma^{a0i}}=-4k\oint{dS_{i}\Sigma^{a0i}}\,.
\end{equation}
This expression is invariant under spatial coordinate transformations and time reparametrization. It is also a vector under SO(3,1) symmetry, thus the zero component of $P^a$, which represents the energy, assumes different values for different observers. The energy-momentum vector (\ref{eq13}) shall be used in the following section to evaluate the total energy-momentum of a spacetime containing a conical defect.

\section{Conical defects}\label{conical}

A general conical defect depends on three constant parameters and may be written as \cite{tod1994conical}
\begin{equation}\nonumber
ds^{2}=-(dt+\alpha d\phi)^{2}+dr^{2}+\beta^{2}r^{2}d\phi^{2}+(dz+\gamma d\phi)^{2}\,.
\end{equation}
The parameter $\alpha$ is related to the rotational characteristic of the defect and, in this paper, will be considered zero, so the geodesic equations are parametrized with relation to the time $t$ in subsection \ref{geodesic}. Therefore, considering $\alpha=0$ in the above equation, we arrive at
\begin{equation}\label{eq14}
ds^{2}=-dt^{2}+dr^{2}+(\beta^{2}r^{2}+\gamma^{2})d\phi^{2}+dz^{2}+2\gamma dzd\phi\,,
\end{equation}
where the coordinate $\phi$ is continuous at $2\pi$. The parameter $\gamma$ is related to a dislocation and the parameter $\beta$ is related to a disclination. This metric is the asymptotic metric of a cylindrical symmetric gravitational wave over a large radial distance \cite{tod1994conical,berger1995asymptotically}. By making $\gamma=0$, we retrieve the metric of a cosmic string.

In context of geometry, the defects may be modeled from step functions as
\begin{equation}\label{eq15}
\gamma(r)=
\begin{cases} 
      \gamma_{0}\,, & r>a\,, \\
      0\,, & r\leq a\,,
\end{cases}
\end{equation}
and
\begin{equation}\label{eq16}
\beta(r)=
\begin{cases} 
      \beta_{0}\,, & r>a\,, \\
      1\,, & r\leq a\,,
\end{cases}
\end{equation}
where the defect is obtained by making $a\rightarrow 0$. It is worth noticing that we recover Minkowski's spacetime when $\gamma=0$ and $\beta=1$. The curvature tensor is zero everywhere outside the defect, but not the torsion tensor.
In order to evaluate the torsion tensor, we need to establish a reference frame adapted to a spatially static observer in this spacetime, i.e., $u^{\mu}=e_{(0)}\,^{\mu}=(u^{0},0,0,0)$. This condition results in the observer measuring the gravitational effect without the interference of his kinematic effects. A tetrad field associated with the line element (\ref{eq14}) and adapted to a spatially static observer is
\begin{equation}\label{eq17}
e_{a\mu}=\left(
\begin{array}{cccc}
 -1 & 0 & 0 & 0 \\
 0 & \cos (\phi ) & -r \sin (\phi ) \beta (r) & 0 \\
 0 & \sin (\phi ) & r \cos (\phi ) \beta (r) & 0 \\
 0 & 0 & \gamma (r) & 1 \\
\end{array}
\right)\,.
\end{equation}
The non-null torsion tensor components (\ref{eq4}) constructed from the tetrads (\ref{eq17}) are
\begin{align}
&T^{(1)}\,_{21}=\sin{\phi}\left(1-\beta-r\beta'\right)\,,\label{eq18}\\
&T^{(2)}\,_{21}=-\cos{\phi}\left(1-\beta-r\beta'\right)\,,\label{eq19}\\
&T^{(3)}\,_{21}=\gamma'\label{eq20}\,.
\end{align}
Outside the source $r>a$, we have $\beta'=\gamma'=0$, but the torsion tensor still has two non-zero components when $\beta\neq 1$, namely $T^{(1)}\,_{21}$ and $T^{(2)}\,_{21}$. 

The following subsections are devoted to measuring the intensity and direction of the dislocation, evaluating the gravitational energy and the geodesic equations of particles in the presence of a conical defect.

\subsection{Burgers vector}\label{burgers}

An otherwise closed loop in the flat Minkowski spacetime will fail to close in the presence of a dislocation, i.e., a distortion caused by a translation. This closure failure of the loop may be used to measure the dislocation intensity and direction. The torsion is the dislocation density, thus by integrating the torsion into a surface $S$ that encompasses the defect, we obtain the defect intensity
\begin{equation}\label{eq21}
b^{(i)}=\frac{1}{2}\int_{S}{T^{(i)}\,_{jk} dx^{j}\wedge dx^{k}}=\oint_{\mathcal{C}}{e^{(i)}\,_{j}dx^{j}}\,,
\end{equation}
where $\wedge$ indicates the exterior product and the curve $\mathcal{C}$ is the closed curve that bounds the surface $S$. This vector has a dimension of length and in solid state is called Burgers vector. This vector will be assumed to be the multiple of an integer in section \ref{results}.
Using the tetrads (\ref{eq17}), the Burgers vector may be calculated trivially by choosing the contour as an orthogonal circle around the longitudinal $z$ axis, obtaining
\begin{equation}\label{eq22}
b^{(i)}=\int_{0}^{2\pi}{e^{(i)}\,_{2}d\phi}=(0,0,2\pi\gamma_{0})\,.
\end{equation}
Hence, the conical defect contains a dislocation on the $z$ axis with an intensity proportional to $\gamma$. The disclination intensity is measured by Frank's matrices and is not the main concern of this article. For a discussion of the Frank's matrices, see Refs. \cite{puntigam1997volterra,kleman2008disclinations}.

\subsection{Gravitational energy}\label{energy}

The total gravitational energy of the spacetime can be evaluated directly from equation (\ref{eq13}).
After projecting the torsions (\ref{eq18}-\ref{eq20}) into the tangent space, we obtain the non-vanishing $\Sigma^{(0)bc}$ components as
\begin{align}
&\Sigma^{(0)(0)(1)}=\cos{\phi}\left(\frac{1-\beta-r\beta'}{2r\beta}\right)\,,\label{eq23}\\
&\Sigma^{(0)(0)(2)}=\sin{\phi}\left(\frac{1-\beta-r\beta'}{2r\beta}\right)\,,\label{eq24}
\end{align}
thus,
\begin{equation}\label{eq25}
\Sigma^{(0)01}=e_{b}\,^{0}e_{c}\,^{1}\Sigma^{(0)bc}=\frac{1-\beta-r\beta'}{2r\beta}\,.
\end{equation}
The determinant is $e=r\beta$, therefore the total energy is
\begin{equation}\label{eq26}
\begin{split}
E=P^{(0)}&=\frac{1}{4\pi}\int_{V}{\partial_{r}\left(e\Sigma^{(0)01}\right)}\\
&=\frac{1}{8\pi}\oint_{S}{\left(1-\beta-r\beta'\right)}\\
&=\left(\frac{1-\beta_{0}}{4}\right)L\,,
\end{split}
\end{equation}
where $L$ is the defect length. The energy per unit of length,
\begin{equation}\label{eq27}
\epsilon\equiv\frac{E}{L}=\frac{1-\beta_{0}}{4}\,,
\end{equation}
is finite. When the disclination disappears, i.e., $\beta_{0}=1$, the total energy goes to zero. The result (\ref{eq26}) is the same of that of a pure cosmic string obtained in Ref. \cite{maluf2001space}. In the same reference the energy for a pure dislocation ($\gamma_{0}\neq0$, $\beta_{0}=1$) is found to be zero, therefore we may conclude that the energy of the two combined defects is the sum of the energy of the individual defects.

\subsection{Geodesics}\label{geodesic}

\begin{figure}[htbp]
\begin{minipage}[t]{0.45\textwidth}
\centering
			\includegraphics[width=0.8\textwidth]{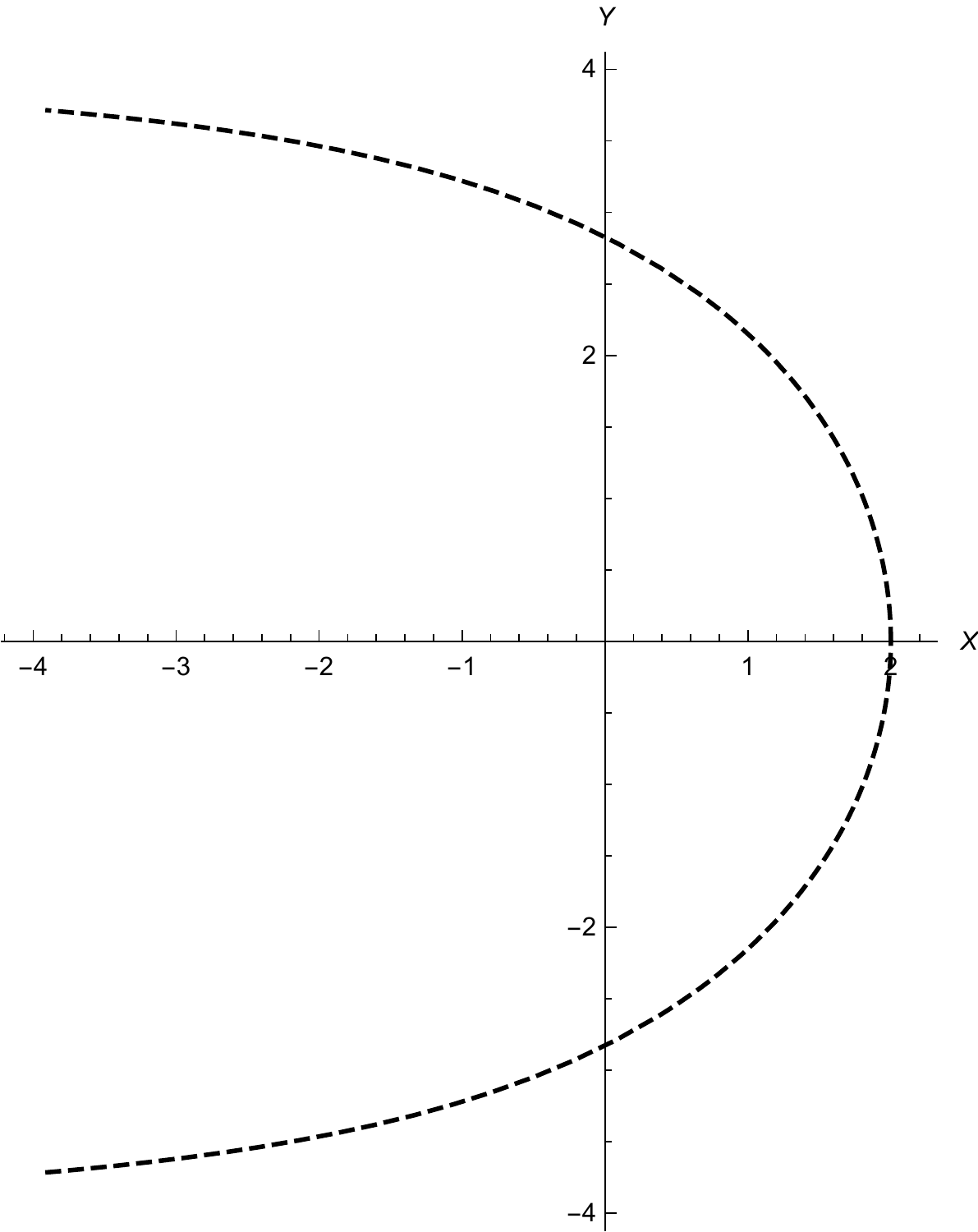}
		\caption{2-dimensional parabolic orbit for $\beta_{0}=1/2$.}
		\label{fig1}
\end{minipage}
\qquad
\begin{minipage}[t]{0.45\textwidth}
\centering
			\includegraphics[width=0.95\textwidth]{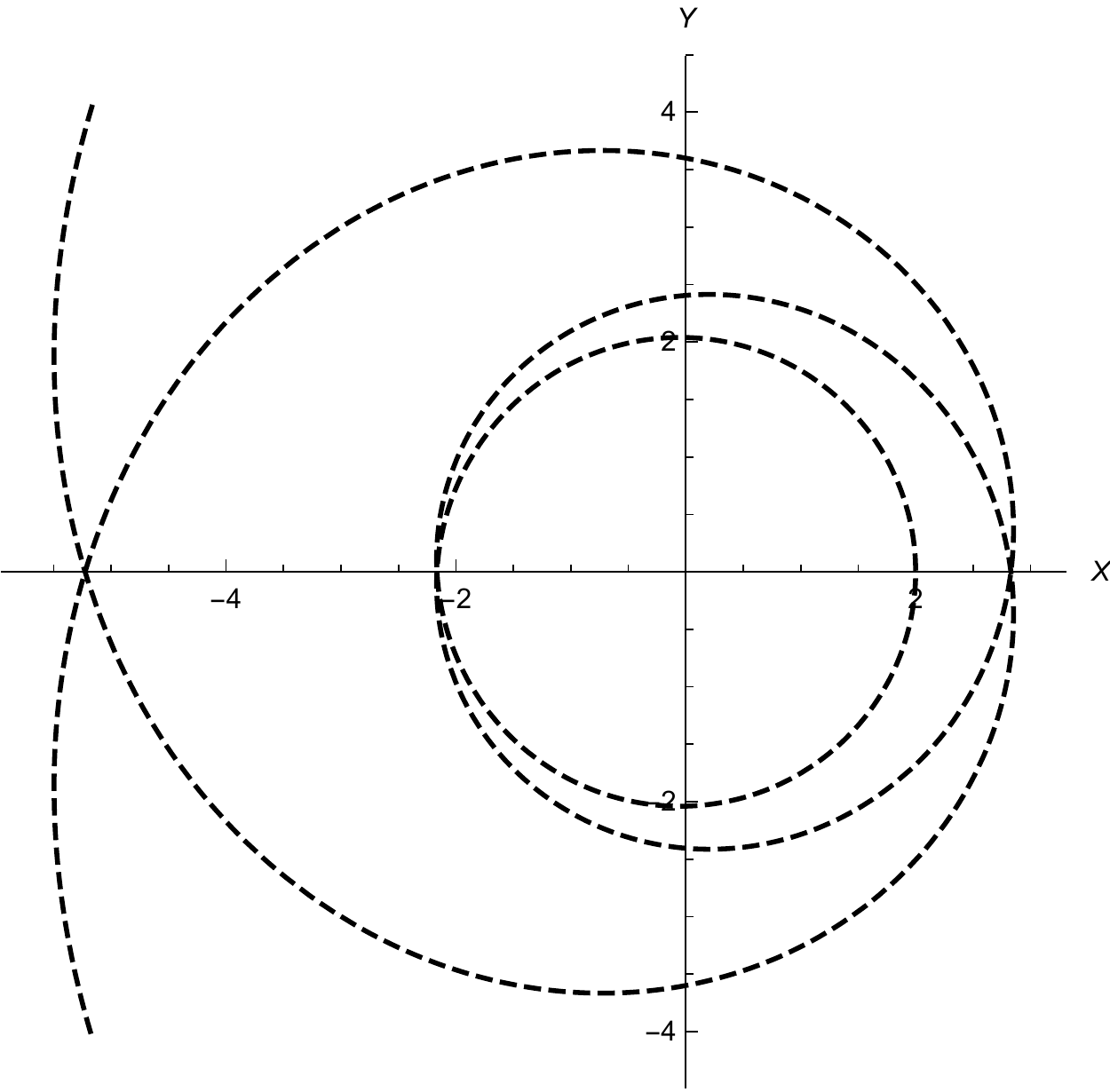}
		\caption{2-dimensional closed orbit $\beta_{0}=1/8$.}
		\label{fig2}
\end{minipage}
\end{figure}

\begin{figure}[htbp]
\centering
			\includegraphics[width=0.3\textwidth]{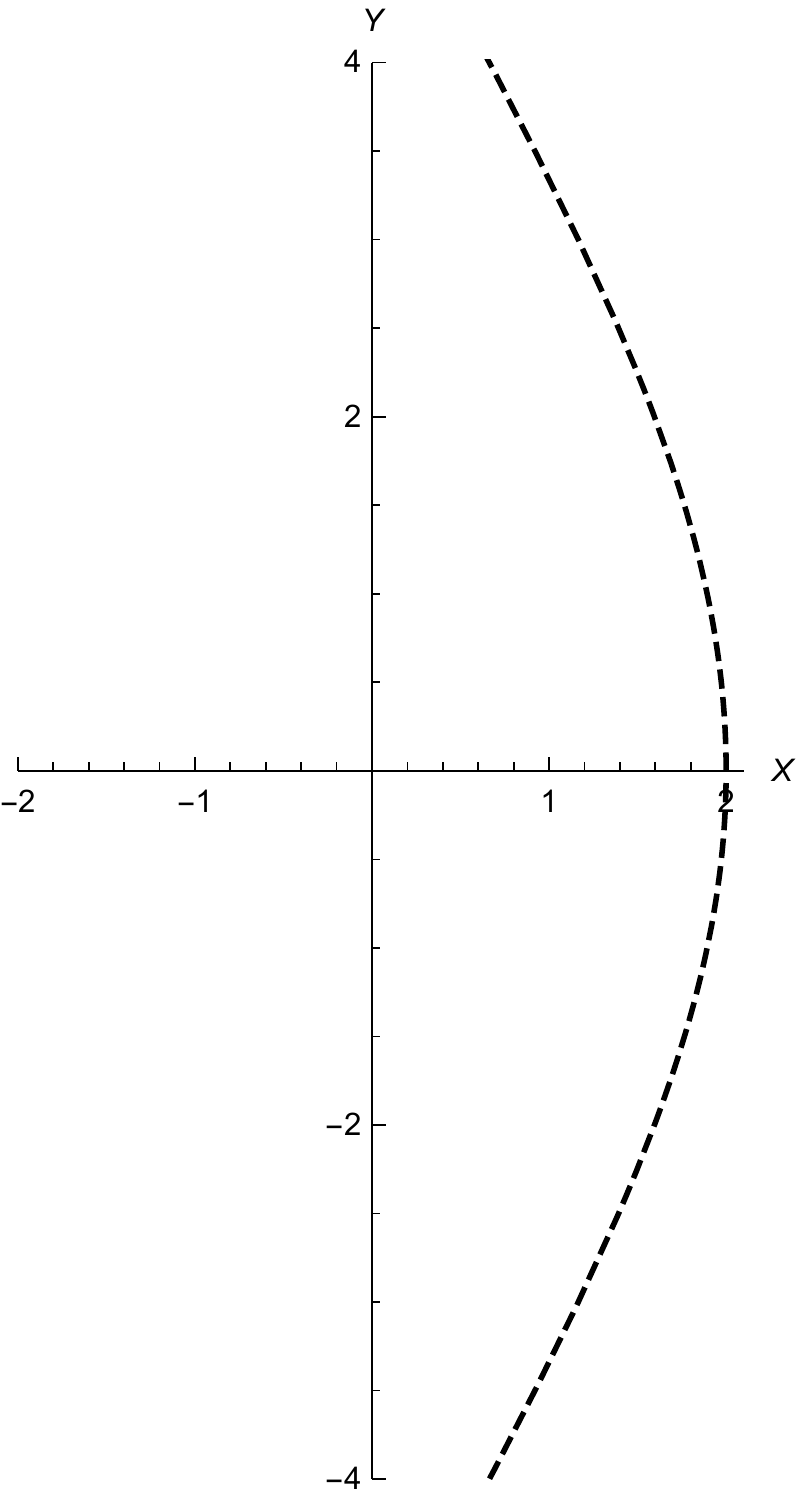}
		\caption{2-dimensional hyperbolic orbit $\beta_{0}=3/4$.}
		\label{fig3}
\end{figure}

The non-vanishing Christoffel symbols ${}^0{\Gamma}^{\lambda}\,_{\mu\nu}$ for the conical defect (\ref{eq14}) are
\begin{equation}\nonumber
{}^0{\Gamma}^{1}\,_{22}=-r\beta_{0}^{2}\,,\qquad\qquad
{}^0{\Gamma}^{2}\,_{12}=\frac{1}{r}\,,\qquad\qquad
{}^0{\Gamma}^{3}\,_{12}=-\frac{\gamma}{r}\,,
\end{equation}
thus, the geodesic equations are
\begin{align}
&\ddot{t}=0\,,\label{eq28}\\
&\ddot{r}-r\beta_{0}^{2}\dot{\phi}^{2}=0\Rightarrow \ddot{r}-l^{2}\beta_{0}^{2}r^{-3}=0\,,\label{eq29}\\
&\ddot{\phi}+2r^{-1}\dot{r}\dot{\phi}=0\Rightarrow \dot{l}=0\,,\label{eq30}\\
&\ddot{z}-2\gamma r^{-1}\dot{r}\dot{\phi}=0\Rightarrow\dot{z}+\gamma\dot{\phi}=constant \,,\label{eq31}
\end{align}
where $l=r^{2}\dot{\phi}$ is the angular momentum of the particle per unit of mass and the dots indicate derivatives with respect to the proper time.
For the line element (\ref{eq14}), the angular momentum of the particle is conserved and the proper time coincides with the coordinate time $t$. The dislocation affects the particle longitudinal motion, since if $\gamma\neq0\Rightarrow \dot{z}\neq constant$ and the disclination $\beta_{0}<1$ may result in a periodic motion. 
The coupled equations (\ref{eq28}-\ref{eq31}) can be analytically solved with the initial conditions $r(0)=r_{0}$, $\dot{r}(0)=\dot{r}_{0}$,  $\phi(0)=\phi_{0}$, $z(0)=z_{0}$, $\dot{z}(0)=\dot{z}_{0}$. Thus we obtain
\begin{align}
r(t)&=\frac{1}{r_{0}}\sqrt{\left(r_{0}\dot{r}_{0}t+r_{0}^{2}\right)^{2}+l^{2}\beta_{0}^{2}t^{2}}\,,\label{eq32}\\
\Delta\phi &=\phi-\phi_{0}=\frac{1}{\beta_{0}}\arctan{\left(\frac{l\beta_{0} t}{r_{0}^{2}+r_{0}\dot{r}_{0}t}\right)}\,,\label{eq33}\\
z(t)&=z_{0}+\left(\dot{z}_{0}+\frac{l\gamma}{r_{0}^{2}}\right)t-\gamma\Delta\phi\,.\label{eq34}
\end{align}
The existence of a disclination, i.e., $\beta_{0}<1$, implies the possibility of periodic motion over a time period that depends on the disclination intensity. This fact can be perceived by inverting the equation (\ref{eq33}) to
\begin{equation}\label{eq35}
t=\frac{r_{0}^{2}\tan{\left[\beta_{0}\left(\phi-\phi_{0}\right)\right]}}{l\beta_{0}-r_{0}\dot{r}_{0}\tan{\left[\beta_{0}\left(\phi-\phi_{0}\right)\right]}}\,.
\end{equation}
The limit $t\rightarrow\infty$, together with $\dot{r}_{0}=0$, in equation (\ref{eq33}) lead to $\beta_{0}\Delta\phi=\pi/2$. If $\beta_{0}=1$ the particle follows a straight line $\Delta\phi=\pi/2$. However, if $\beta_{0}<1$ the particle will follow a non-straight geodesic and perform a closed orbit $\Delta\phi=2\pi$ if $\beta_{0}=1/4$. By defining an angle 
\begin{equation}\label{eq36}
\theta\equiv\beta_{0}\Delta\phi\,,
\end{equation}
the orbit behavior may be analyzed for each $\beta_{0}$ by mapping the maximum $\Delta\phi$ that provides $\theta=\pi/2$. Thus, we may have parabolic, closed and hyperbolic orbits, as shown in Figures \ref{fig1}, \ref{fig2} and \ref{fig3}, respectively, for the initial conditions (in natural units)
\begin{equation}\label{eq37}
r_{0}=2\,,\qquad l=1\,,\qquad \phi_{0}=0\,,\qquad\dot{r}_{0}=z_{0}=\dot{z}_{0}=0\,.
\end{equation}

Closed orbits in the presence of conical defects are always limited in time, thus in these cases the particle escapes with an offset angle $\Delta\Phi=\phi(t=\infty)-\phi_{0}$ at the exit given by the limit $t\rightarrow\infty$ in the equation (\ref{eq33}), yielding
\begin{equation}\label{eq38}
\Delta\Phi\equiv\lim_{t\rightarrow\infty}{\Delta\phi}=\frac{1}{\beta_{0}}\arctan{\left[\frac{l\beta_{0}}{r_{0}\dot{r}_{0}}\right]}\,.
\end{equation}
From equation (\ref{eq38}) we may conclude that when $\dot{r}_{0}=0$ and $l\neq0$, the offset angle is always $\frac{\pi}{2\beta_{0}}$.
Therefore, a disclination intensity of a conical defect can be measured by releasing a particle, with specific initial conditions, close to the defect and measuring its offset angle $\Delta\Phi$ after a long period of time.

The number of loops that the particle performs may be obtained from equation (\ref{eq33}). In the absence of a conical defect, a particle released with $\phi(t=0)=0$ will escape at an angle $\phi(t\rightarrow\infty)=\pi/2$, performing $1/4$ of a loop. Therefore, in the presence of a defect it is more convenient to count the additional $N$ loops to $1/4$. It should be noted that $N$ is not an integer a priori. In the limit
\begin{equation}\nonumber
\lim_{t\to\infty}\Delta\phi=2\pi\left(N+1/4\right)
\end{equation}
the number of additional loops $N$ may be found as
\begin{equation}\label{eq39}
N=\frac{1}{2\pi\beta_{0}}\arctan{\left(\frac{l\beta_{0}}{r_{0}\dot{r}_{0}}\right)}-\frac{1}{4}\,.
\end{equation}
Therefore, the number of loops performed by the particle depends on the disclination intensity $\beta$, the initial radial position $r_{0}$, the initial radial velocity $\dot{r}_{0}$ and the angular momentum of the particle per unit of mass $l$.
For a particle with zero initial radial velocity, the total number of loops depends only on the dislocation intensity since
\begin{equation}\label{eq40}
N=\frac{1}{4}\left(\frac{1-\beta_{0}}{\beta_{0}}\right)
\end{equation}
for $\dot{r}_{0}=0$. In the absence of topological defects, i.e., $\beta_{0}=1$, the particle does not perform additional loops as it is expected.
Besides the orbit of particles around conical defects be always finite in time, a large number of loops may be obtained by making $\beta_{0}$ sufficiently small.

The longitudinal motion of a particle occurs in the presence of a dislocation. Thus, combined with the circular motion caused by the disclination, the dislocation engender a helical motion for the particle. The evaluation of the helices steps grants the possibility to measure the dislocation intensity $\gamma_{0}$.
In the following section, we consider the three-dimensional motion described by a particle in the presence of a conical defect with a discrete dislocation intensity.

\section{Geodesic effects of a discrete Burgers vector}\label{results}

\begin{figure}[htbp]
\centering
\begin{minipage}[t]{0.45\textwidth}
\centering
			\includegraphics[width=0.9\textwidth]{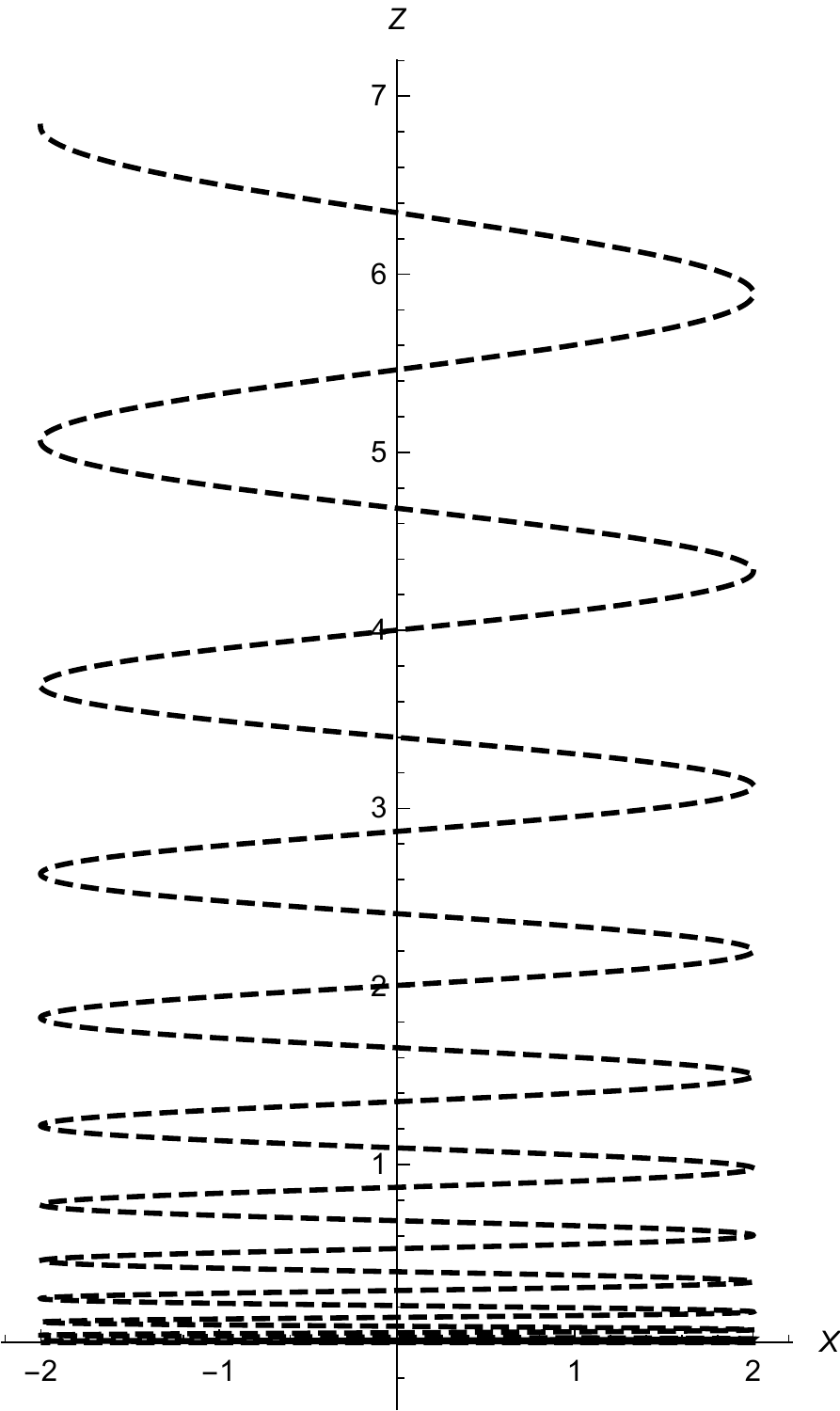}
		\caption{Trajectory in the $XZ$ plane for $\beta=1/100$, $b_{0}=1$ and $n'=1$.}
		\label{fig4}
\end{minipage}
\qquad
\begin{minipage}[t]{0.45\textwidth}
\centering
			\includegraphics[width=0.9\textwidth]{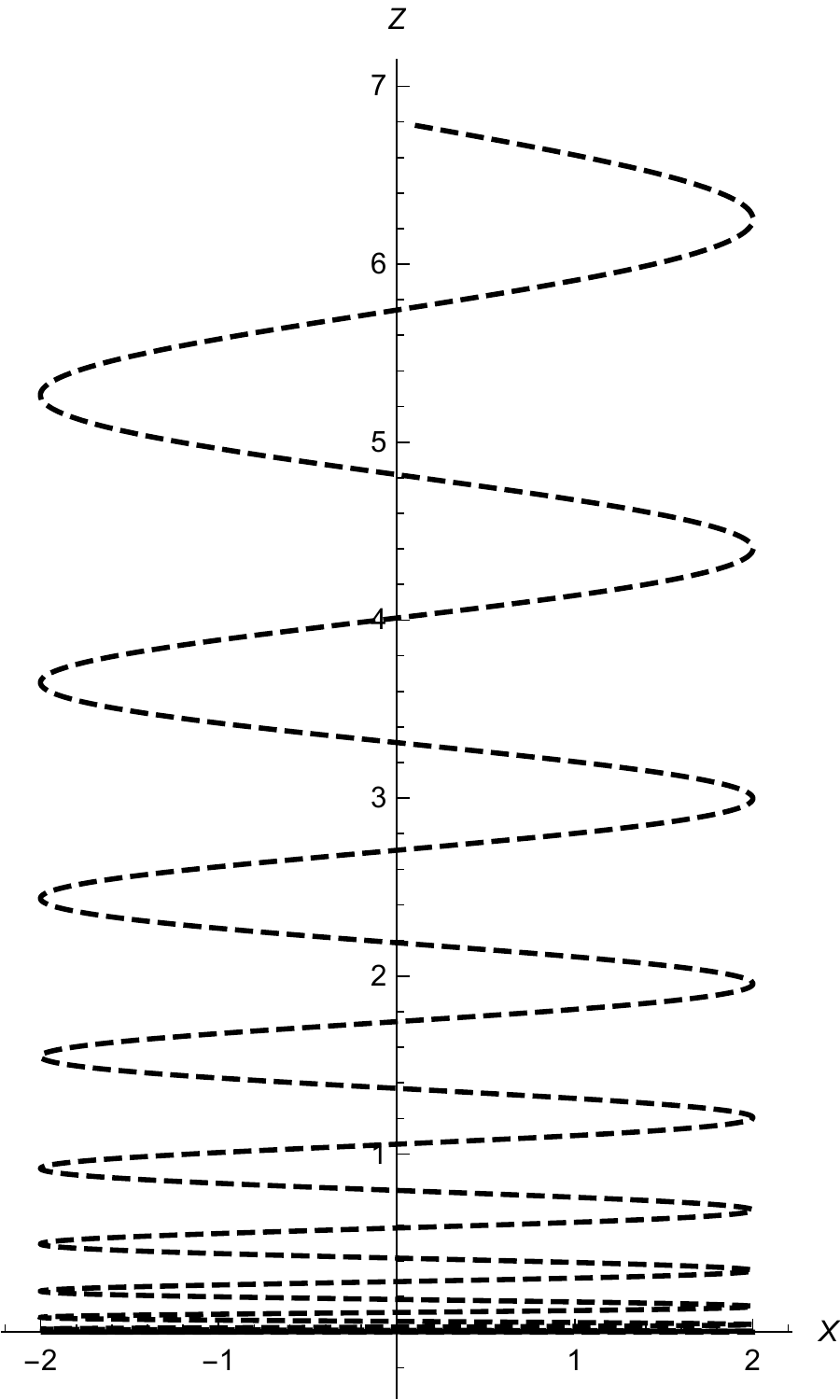}
		\caption{Trajectory in the $XZ$ plane for $\beta=1/100$, $b_{0}=1$ and $n'=4$.}
		\label{fig5}
\end{minipage}
\end{figure}

\begin{figure}[htbp]
\centering
		\includegraphics[width=0.7\textwidth]{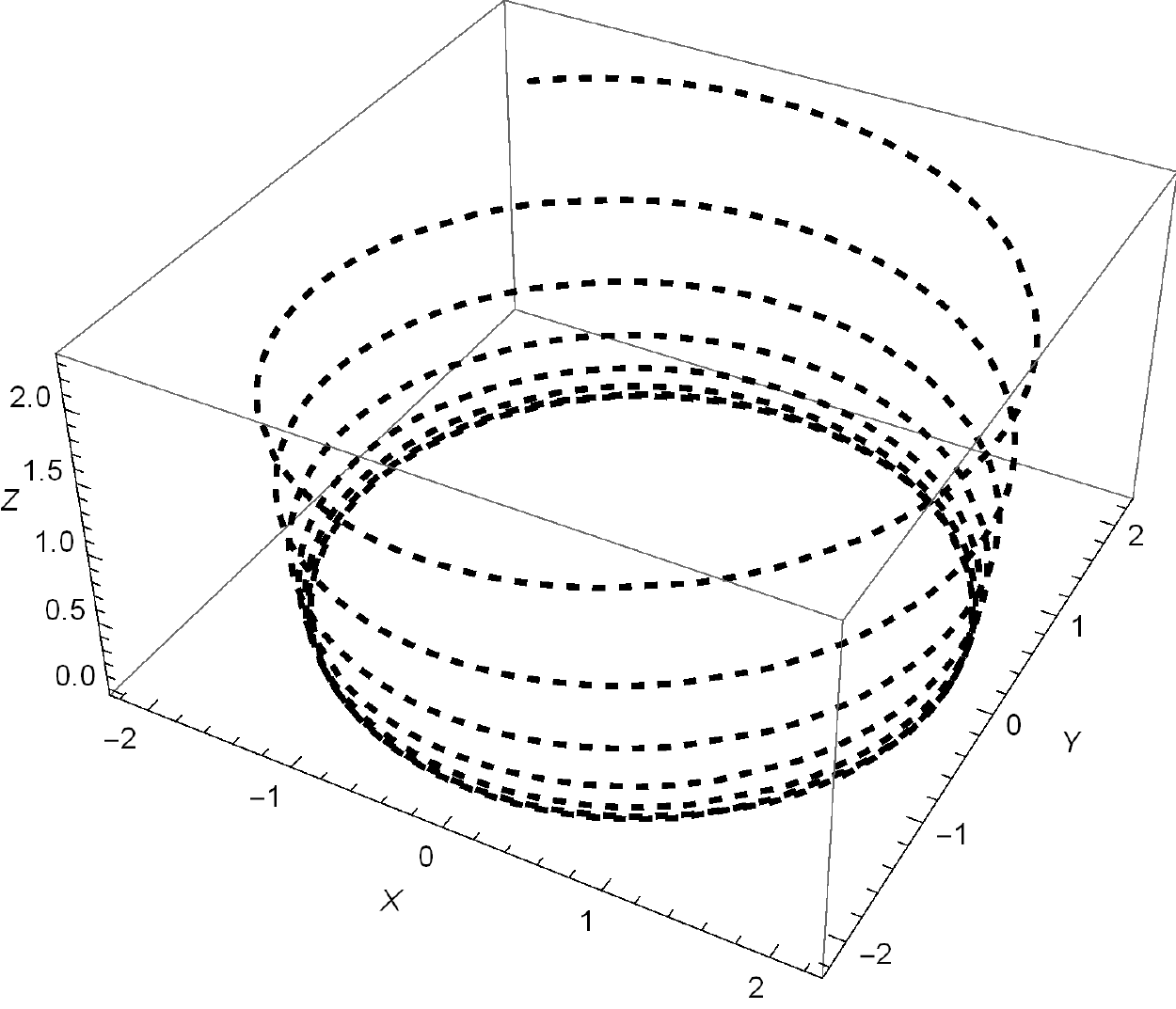}
		\caption{Three-dimensional trajectory for $b_{0}=1$, $n=4$ and $\beta=1/100$.}
		\label{fig6}
\end{figure}

The concept of discrete dislocation intensity was introduced with the aim of bringing Plank's constant into gravitational theories in the pioneering works of Magnon \cite{magnon1991spin}, Ross \cite{ross1989planck}, Sabbata and collaborators \cite{sabbata1991quantum}. As seen in the subsection \ref{burgers}, the dislocation intensity is measured by a vector called Burgers vector. 
This vector has a dimension of length, therefore the Plank's constant was introduced postulating that this vector was an integer multiple of the Plank length. In this section we assume the same hypothesis and verify the consequences of this assumption on the visible parameters of the geodesic motion of a particle in the presence of the conical defect (\ref{eq14}).
By imposing $b=nb_{0}$, where $n$ is an integer and $b_{0}$ is the Plank length in natural units, we arrive at
\begin{equation}\label{eq41}
\gamma_{0}=\frac{nb_{0}}{2\pi}\,,
\end{equation}
where $\gamma_{0}=\frac{b_{0}}{2\pi}$ denotes the minimum possible intensity for a dislocation. It should be noted that $\gamma_{0}=\left(\gamma_{0}\right)_{n}\,$, while $b_{0}$ is a constant.
This discrete intensity implies a discrete behavior in the motion of a particle. If $b=b_{n}\Rightarrow \gamma=\gamma_{n}$, then
\begin{align}
z(t)=z_{n}(t)&=z_{0}+\dot{z}_{0}t+\left(\frac{l}{r_{0}^{2}}t-\Delta{\phi}\right)\gamma_{n}\nonumber\\
&=z_{0}+\dot{z}_{0}t+\left(\frac{l}{r_{0}^{2}}t-\Delta{\phi}\right)\frac{b_{0}}{2\pi}n\label{eq42}\,.
\end{align}
With the above expression, it is possible to evaluate the distance between a $ n '$ step and the immediately preceding step $ n'-1 $, i.e., $\Delta z_{n,n'}\equiv z_{n,n'}-z_{n,n'-1}$. In order to simplify the result, we will restrict the analysis to the case $\dot{r}_{0}=0$, where
\begin{equation}\label{eq43}
t_{n'}=\frac{r_{0}^{2}}{l\beta_{0}}\tan{\left(2\pi n'\beta_{0}\right)}\,.
\end{equation}
Therefore,
\begin{align}
\Delta z_{n,n'}(t_{n'})&=\left[z_{0}+\dot{z}_{0}+\left(\frac{l}{r_{0}^{2}}t_{n'}-2\pi n'\right)\gamma_{n}\right]\nonumber\\
&-\left[z_{0}+\dot{z}_{0}+\left(\frac{l}{r_{0}^{2}}t_{n'-1}-2\pi (n'-1)\right)\gamma_{n}\right]\nonumber\\
&=\left[\frac{1}{\beta_0}\left\{\tan{\left(2\pi n'\beta_0\right)}-\tan{\left[2\pi (n'-1)\beta_0\right]}\right\}-2\pi\right]\frac{b_{0}}{2\pi}n\,.\label{eq44}
\end{align}
The distance between the steps increases with the number of loops $n'$, but it is always an integer multiple of the Plank's constant.
If the Burgers vector is discrete, there is a minimum discrete distance between two loops that depends on the intensity of the dislocation. The distance between the loops increases with the number of steps $n'$, as shown in Figures \ref{fig4} and \ref{fig5} for the initial conditions (\ref{eq37}).
The three-dimensional trajectory is helical as displayed in Figure \ref{fig6}. 
The intensity of dislocation of a conical defect can then be determined by measuring the steps of the helical motion of a particle released in the vicinity of that defect. Therewith, the excitation number $n$ can be determined if one knows the offset angle, hence $\beta$, and the initial conditions of the particle.

The presence of a dislocation $\gamma$ directly affects the particle longitudinal motion, even when $\dot{z}_{0}=0$.
For a particle with initial conditions $z_{0}=\dot{z}_{0}=0$, the longitudinal velocity of the particle at infinity can be obtained by differentiating the equation (\ref{eq42}) and evaluating the limit $t\rightarrow\infty$, thus obtaining
\begin{equation}\label{eq45}
\dot{z}(t\rightarrow\infty)=\frac{l}{r_{0}^{2}}\frac{b_{0}}{2\pi}n=\frac{b_{0}}{2\pi}\dot{\phi_{0}}n\,.
\end{equation}
The presence of a non-null $\gamma$ causes a longitudinal memory effect on a particle that depends only on the particle initial angular velocity $\dot{\phi}_{0}$ and on the excitation $n$ of the defect.
Therefore, if the Burgers vector is an integer multiple of Plank length, the possible longitudinal velocities for a particle are discrete when the defect no longer acts on the particle's trajectory.
An interesting fact can be observed by assessing the classical kinetic energy of the particle. Initially, when $t=0$, we have
\begin{equation}\label{eq46}
K_{i}=\frac{1}{2}\left(\dot{r}_{0}^{2}+r_{0}^{2}\dot{\phi}_{0}^{2}+\dot{z}_{0}^{2}\right)=\frac{1}{2}r_{0}^{2}\dot{\phi}_{0}^{2}
\end{equation}
and at infinity we have
\begin{align}
K_{f}&=\frac{1}{2}\left(\dot{r}({\infty})^{2}+r({\infty})^{2}\dot{\phi}({\infty})^{2}+\dot{z}({\infty})^{2}\right)\nonumber\\
&=\frac{1}{2}\frac{l^{2}\beta^{2}}{r_{0}^{2}}+\frac{b_{0}^{2}\dot{\phi}_{0}^{2}}{8\pi^{2}}n^{2}\label{eq47}\,.
\end{align}
Thus the variation of the kinetic energy is
\begin{equation}\label{eq48}
\Delta K =\frac{1}{2}\dot{\phi}^{2}r_{0}^{2}\left(\beta^{2}-1\right)+\frac{b_{0}^{2}\dot{\phi}_{0}^{2}}{8\pi^{2}}n^{2}\,.
\end{equation}

The disclination of a conical defect always removes energy from the particle (since $\beta^2-1 <0$) and  the dislocation always provides energy to the particle. One of the most interesting results is that the particle extracts a discrete amount of energy of the defect. 
The possibility that the energy (and also angular momentum) acquired by the particle comes from the gravitational field had already been contemplated in the case of a particle interacting with a gravitational wave \cite{maluf2018variations,carneiro2019energy}.
This possibility implies the existence of a discrete energy for the gravitational field of the defect and, consequently, a discrete disclination intensity. 
\section{Final considerations}\label{conclusions}

This article is devoted to the comprehension of the idea of a discrete Burgers vector by analyzing the geodesic motion of a particle around a conical defect, which can be a possible visible parameter in experimental observations. The Lagrangian formulation of the TERG was revised in section \ref{terg}, from which we extracted the total energy-momentum vector (gravitational field plus the defect) in the form of the expression (\ref{eq13}). 
In section \ref{conical}, we introduced the conical defect (\ref{eq14}) that was considered in this paper and described the gravitation effect of the defect through the spacetime torsion, which produces regular expressions. Thereafter, the closure failure of a loop was calculated at subsection {\ref{burgers}} and we evaluated the energy-momentum for the defect in subsection \ref{energy}. This procedure yielded the same result as for a cosmic string, i.e., the total energy per unit length of a conical defect is the linear combination of the energy of its dislocation $E=0$ and the energy of its disclination $E=(1-\beta)L/4$.
The geodesic equations of a particle in the presence of the defect were solved in subsection \ref{geodesic} and an observational parameter (\ref{eq38}) was found to measure the intensity of the disclination. The orbits of the particles are similar to those of a Newtonian central force, but with the difference that the closed orbits are time-limited, i.e., the particle always escape the defect after a finite time. 
Another interesting observational parameter related to the disclination intensity is the total number of loops performed by a particle in a closed orbit before escaping. Considering the energy density (\ref{eq27}) and the number of additional loops (\ref{eq40}), we obtain
\begin{equation}\label{eq49}
\epsilon=\beta N\,.
\end{equation}
Therefore, by measuring the offset angle, hence $\beta$, and the number of loops, one may obtain the energy density of a conical defect.

The effects of the disclination on the particle motion was considered in section \ref{results} from the imposition of a discrete Burgers vector (\ref{eq41}). The result was a helical motion with a discrete distance between each loop. This discretization appears in the longitudinal velocity and consequently in the kinetic energy of the particle at infinity, thus the energy of the particle varies with respect to a discrete value which is the square of the defect length $nb_{0}$.

The discrete variation of the particle's longitudinal kinetic energy (second term on the right hand side of (\ref{eq48})) may have two explanations: \textit{(i)} the particle extracts energy from the defect, therefore the energy of the defect must be discrete, i.e., $\epsilon=\epsilon_{n}\Rightarrow\beta=\beta_{n}$; or (\textit{ii}) the defect removes energy from the perpendicular direction of movement (first term on the right hand side of (\ref{eq48})) and transfers to the longitudinal motion, also implying $\beta=\beta_{n}$.
Thus, in both cases it seems that by imposing a discrete parameter on the dislocation, the same discrete parameter naturally appears on the disclination. Therefore, the relation
\begin{equation}\label{eq50}
\beta_{0}=1-2\pi\frac{\gamma_{0}}{\Gamma}\Leftrightarrow\gamma_{0}=\frac{\Gamma}{2\pi}\left(1-\beta_{0}\right)
\end{equation}
where $\Gamma$ is a parameter with dimension of length, should be verified. In this case, the disclination intensity may also be written in terms of the Burgers vector length as
\begin{equation}\label{eq51}
\beta_{0}=1-n\frac{b_{0}}{\Gamma}\,,
\end{equation}
thus the energy density becomes
\begin{equation}\label{eq52}
\epsilon_{n}=\frac{b_{0}}{4\Gamma}n\,,
\end{equation}
i.e., a discrete energy per unit of length. This fact may have consequences on the number of loops performed by the particle as equation (\ref{eq40}) becomes
\begin{equation}\label{eq53}
N=\frac{1}{4}\left(\frac{n}{\Gamma/b_{0}-n}\right)\,.
\end{equation}
Therefore, the hypothesis $\gamma_{n}\rightarrow\beta_{n}$ may be verified by analyzing a data set of particles and searching for a discrete parameter $n$ in the form of the equation (\ref{eq53}).

Despite the inherent experimental difficulties related to topological defects, the quantization of Burgers vector provides an observable parameter.
There are several attempts to construct a quantum theory of gravity, such as the Loop quantum gravity (see Ref. \cite{rovelli2008loop} and references there in), String theory (see Ref. \cite{blau2009string} and references there in), Weyl quantization (see Ref. \cite{ulhoa2017quantization} and references there in), but the idea of bringing the quantization through topological defects is a simple and elegant manner, addressing the geometrical origin of the Plank's constant as pointed out by Ross \cite{ross1989planck}.

The possibility of having an observational parameter like the geodesic motion can be of great importance for the detection of topological defects and the possibility of a discrete intensity for these defects. There are speculations about the existence of cosmic strings and other topological defects, and several articles have recently dealt with the effects and interpretations of topological defects \cite{hosseinpour2019dirac,wang2019discussion,flachi2019cosmic,millette2018bosons,kleinert2011new}.
One possibility is that topological defects have acted in the formation of structures in our universe \cite{brandenberger1994topological,durrer2002cosmic}, and the presence of these defects may influence some large-scale behaviors, such as cosmic ray energy \cite{protheroe1996limits} and the structure of galaxies \cite{thomas2009rotation,lemos1991topological}. Therefore, large structures in our universe, such as galaxies, may provided experimental parameters for the analysis proposed in this paper, in particular the possibility that the closure failure of a loop around the defect is a multiple of the Plank length.

\end{document}